# Education for All: Remote testing system with gesture recognition and recording

Rivindu Perera
Department of Computer Science
Informatics Institute of Technology
Colombo, Sri Lanka
*pererarivindu@hotmail.com*

*Abstract*—Etymologically, in Latin expresses educare, that means to bring out, or be engaged in the infinite process of learning to present to the society as a valuable citizen. However, unfortunately especially in third world countries, education cannot be achieved due to, lack of inorganic and organic resources. However, many third world countries have embraced the concepts such as One Laptop per Child, facilitating the students to learn. The effective adaptation of these concepts has being launched through many government and non-government projects, providing inorganic resources. However, inorganic resources alone cannot provide quality education, as learning needs assessment procedures, feedback generators and trainers who could guide the students to gain knowledge.

This paper attempts to introduce an acceptable solution that can be used to address facilitating resources to enhance the learning experience through enabling organic resources such as teachers, instructors and trainers on a remote mode through technology. This paper introduces a software system that is used to design and distribute examinations and detect gestures of students while answering remotely. The feature enables the teacher or instructor to gain a better understanding about the learner's attitude when taking the assessment.

The content of the paper is organized to give the basic idea of the system and it includes description of the system and practical effectiveness of the system with evaluations from different views.

A java enabled computer with a webcam and internet access is the minimum requirements to be able to use the proposed system. The development platform is based on java, with the use of Chilkat to maintain an asynchronous connection with the FTP server. iGesture and Yuille approach play major role in gesture detection and recognition.

I. INTRODUCTION

In the recent times, blended learning plays a major role in the world of education, with the use of computer based learning systems. However, when considering distance learning approach in remote testing, it is generally impossible to detect gestures, attitude and feelings of student with current methods and technologies that are available. This is mainly due to limited human capabilities and lack of suitable methodologies.

This proposed system is designed to introduce a new methodology that is already implemented as a prototype which can be used to create a remote testing environment. Main idea of this project is to supply a designing interface for teachers to design a test and distribute it over the internet and which can be translated in to an exam in the middle point of its lifecycle. Then students can use the testing software which is part of the main system and can start their exam and interestingly this front-end software can record and will supply comments on gestures of students during the exam.

The system is capable of capturing the emotional gestures of students during the assessment, allowing the lecturer to identify his/her behavior, attitude and feelings through the facial expressions of the student. This system enables creating an assessment environment that represents a real world examination conditions with less resource personnel such as examiners. Additionally, this system can be used as a learning management system that could help students to manage and access their lecture notes and assignments.

## II. TESTING SYSTEM

In this third generation testing system the most important part is the specifically formatted documents. The prototype name the formatted documents as RTS files (remote testing system files) that are capable of going through the internet and enable to play different roles at each point of software system as a compressed file. The system consists of following three major files VQP file, Windows media audio/video file and Encrypted file that are packed inside RTS file.

- VQP file
  This stands for the virtual question paper and consists of questions at the beginning of its life cycle and after answering it contains answers for the questions.
- Windows media Audio/video file
  This file contains the gesture video and recorded voice during the examination.
- Encrypted file
  Consist of audio and video configuration details of student's machine and indicate whether certain options are disabled or not.

### A. *Outline of the architecture and flow of control*

The system is designed according to the layered architecture where presentation and real application logic is kept separately allowing them to be developed individually. Figure 1 shows how the defined architecture is implemented using current technologies.

The flow of the testing system as follows:
  i. Lecturer designs the test using the exam designer and defines the answers for the designed exam and this is called VQP file.
 ii. The system compress the VQP file resulting it to work as a single file called RTS file. During this process system will prompt for the user (lecturer) giving a chance to associate a passkey with the file. This somewhat like a password but exists only for a particular session. (like Bluetooth passkey)
iii. Then the lecturer specifies to which ftp servers the file should be sent to and then system will send the file to those named servers.
 iv. The application interface for student is communicating asynchronously with the ftp server and will inform when the file appeared in the server.
  v. Students will then be able to download the file and if the file is associated with a passkey then the application interface will request for the key. Also at the same time it will check whether audio and video settings are active and will run a test to confirm it.
 vi. Whenever the passkey is entered the exam will start and gesture video will be recorded together with the audio.
vii. Finally after the time period or if the student submit the file, new VPQ file , media file and encrypted file will be transmitted to the lecturer.

### B. *Life cycle of the virtual question paper*

Virtual question paper has three important points in its life cycle:

- ✓ Starting point: Where the VQP file is designed using exam designer of the system by lecturer.
- ✓ Middle point: Where the distribution happens and converts it to a real question paper format.

- ✓ Ending point: Where the file is transferred back to the lecturer but with answers supplied by student.

The main feature of this file is it can be used not only to store multiple choice questions but also for structured type of questions where students can write their own answers. In many testing systems which are currently used today are not capable of handling both of these formats with one interface. But this software system will be able to handle them through MDI formats (Multiple Document Interface formats). The main logic used in this approach heavily dependent on use of the RTS file and its format and encryption. The implementation is done in Java programming language.

### C. Gesture recording and recognition

Normal process of gesture recording and recognition is based on, an image processing method and how data is extracted from the processing lab using a programming language. However, this general way of extraction and recognition is not applicable for a testing system which uses gesture recognition. Hence enhance gesture recognition a combination of Yuille approach and image processing methods are introduced to recognize gestures.

Yuille approach is one of the traditional ways of identifying the static images form a dynamic stream using a template that has been provided as a part of the human body. This method is widely used in many systems for face detection. However, this method was not used with the launch of Matlab processing techniques. A research done by University of California in 2003 clearly defined how many algorithms can be used in combination provide better result in areas like image parsing [1]. It was found through research that Yuille approach can be used to detect motions. Figure 2 shows a template which is used to detect and recognize face gestures using Yuille approach.

Nevertheless, for a testing system similar to the proposed system, it is not at satisfactory level gesture recognition. In addition a survey carried out by University of Alberta with regards to light source detection methods provides clear explanation on the importance of use of appropriate approaches for gesture recognition [2]. The important drawback of the methods such as Yuille approach monitoring will be limited to a particular part of the human body, where as in real environments the total surroundings can be monitored. For example if the provided template is of a human face then with Yuille approach will only record gestures about face. To overcome this difficulty of monitoring the student a gesture recognition framework, iGesture framework has been used with Yuille approach. The selection of this framework used with the testing system provides many other advantages:

- ➢ iGesture allows the programmer to implement his or her own recognition algorithm if the provided algorithms cannot meet with the environment where it is used.
  This feature allows further developments to be carried out for gesture recognition part of the testing system as it works not as a embedded code but as a another layer parallel to the main application.

- ➢ iGesture is distributed as a single jar file that can be used as a library unit. This feature makes this more suitable for a testing system like this, which should be distributed freely for the use of all.

Gesture recording is done in the application which is provided for students which they use as their testing environment. This is carried out by embedding the code needed to record the video stream inside the main application interface and students will not be able to access or do any change to the video or audio recording sub system.

### D. User interfaces and application logic implementation

Testing system is designed with single but multi role interface which will be able to work with student and also with the lecturer based on their account details and the characteristics of the RTS file it received.

User interface is designed in Java language and therefore the system will be cross platform to suit for the different users and also will be able to maintain a better link between other subsystems which are implemented in Java. The common interface mainly plays two major roles in two different ends:

- Interface for the lecturer with exam designer, marking system and gesture recognition sub system.
- Interface for the student with asynchronous FTP client, gesture recording and environment status testing sub system.

Interface for the lecturer is created identifying the required functions that would help the lecturer to design multiple choice questions and also structured questions. When using this system lecturer will be able to select the interface he or she like from two main interfaces provided one with Java and other one is implemented in Qt in order to provide better user experience in designing the test using graphical components such as checkboxes and radio buttons by dragging and dropping to the main design interface. Also gesture recognition system is linked to this interface with implementation as described in section 2.3.

Interface for the students is designed with embedded FTP client which will support it to maintain an asynchronous connection with the FTP server and will inform whenever the file is received. To implement the embedded client Chilkat java ftp library is used and rest of the interface is also designed in Java for better compatibility.

### E. LMS plug-in awareness

As an additional feature this testing system can be extended with a LMS (Learning Management System) as a plug-in. This idea of extending current systems with plug-ins to combine other LMS systems is first introduced by a research done in 2007 about computer-aided learning [5]. For an example the plug-in interface which is designed with the testing system will allow students to access some selected learning management system such as OLAT and Claroline to maintain list of supported lecture materials only when the system is not in its examination mode without active test.

### III. CONCLUSION

This paper described a testing system which use gesture recording and recognition which is designed for student learning in rural areas to overcome the main problem they face, the lack of organic resources. Some key requirements of the system and behavior of the system are mentioned in previous sections providing implementation details where necessary.

This paper evaluated and described the above details taking the prototype which is designed to provide the basic idea of the system. As with the time further work exists in the side of implementing the complete system and also already testing this prototype in-house impressed the author for many improvements of the system like developing the same system as a SAAS application where this can be used from anywhere in the world as a web 2.0 application. Also after the completion of the system it is planned to test this in some rural areas in Sri Lanka and to get feedback on user interface of the system and its effectiveness.

APPENDIX

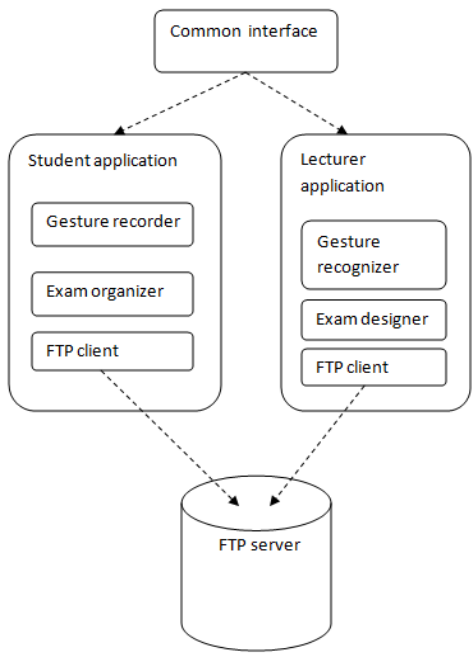

Fig 1: System architecture

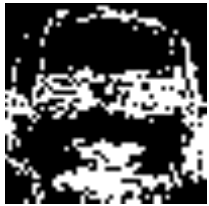

Fig 2: Yuille template